\documentstyle[twocolumn,aps,psfig]{revtex}
\begin{document}
\draft
\flushbottom
\twocolumn[
\hsize\textwidth\columnwidth\hsize\csname @twocolumnfalse\endcsname

\title{ Optical control of photon tunneling through an array of nanometer scale cylindrical
channels }

\author{I. I. Smolyaninov, A. V. Zayats$^{(1)}$, A. Stanishevsky, and C. C. Davis }

\address{  Electrical and Computer Engineering Department,
University of Maryland, College Park, MD 20742}
\address{$^{(1)}$ Department of Pure and Applied Physics,
Queen's University of Belfast, Belfast BT7 1NN,UK}

\date{\today}
\maketitle
\tightenlines
\widetext
\advance\leftskip by 57pt
\advance\rightskip by 57pt

\begin{abstract}
We report first observation of photon tunneling gated by light at
a different wavelength in an artificially created array of
nanometer scale cylindrical channels in a thick gold film.
Polarization properties of gated light provide strong proof of the
enhanced nonlinear optical mixing in nanometric channels involved
in the process. This suggests the possibility of building a new
class of "gated" photon tunneling devices for massive parallel
all-optical signal and image processing.
\end{abstract}


]

\narrowtext

\tightenlines

In our recent work \cite{prl,apl} some of the unusual  properties
of naturally occurring extremely small pinholes in thick gold
films covered with highly optically nonlinear polydiacetylene
layers have been studied. Measurements of photon tunneling through
individual nanometer scale pinholes has provided strong indication
of "photon blockade" effect similar to Coulomb blockade phenomena
observed in single-electron tunneling experiments \cite{prl}. We
have also reported observation of photon tunneling being gated by
light at a different wavelength in similar but somewhat larger
pinholes \cite{apl}. These observations suggest possibility of
building a new class of "gated" photon tunneling devices for
all-optical signal and image processing in classical and quantum
communications and computing. A big step forward would be to
artificially fabricate such devices in controllable and
reproducible manner. Moreover, massive parallel all-optical signal
processing can be achieved if large arrays of such pinholes with
enhanced nonlinear properties can be created. However, transition
from individual pinholes to a periodic array of them leads to new
physics involved in the nonlinear optical processes.

Linear optical properties of metallic films perforated with an
array of periodic subwavelength holes and exhibiting the so-called
extraordinary enhanced optical transmission have been intensively
studied recently (see \cite{ebbesen-98,krishnan-01,prl-01} and
Refs. therein). Although the exact nature of the enhanced
transmission in two-dimensional case (which is significantly
different from a better understood one-dimensional case of
subwavelength slits) is still not fully established, the role of
surface plasmon polaritons (SPPs) in this process is commonly
accepted. SPP Bloch waves can be excited by incident light on the
interfaces of a periodically nanostructured film, which may be
treated as a surface polaritonic crystal
\cite{prl-01,glass-84,barnes-96}. Resonant light tunneling via
states of surface polariton Bloch waves is one of the mechanisms
involved in the enhanced transmission.

A fundamental difference between conventional photonic crystals and surface
polaritonic crystals is a different electromagnetic field
distribution close to the surface. Surface polariton is an
intrinsically two-dimensional excitation whose electromagnetic
field is concentrated at a metal interface. Thus, in contrast to
photonic crystals, strong electromagnetic field enhancement takes
place at a SPP crystal interface related to the surface polariton
field localization \cite{raether}. Such enhancement effects are
absent in photonic crystals. Field enhancement results in much
stronger nonlinear response achieved with surface polaritonic
crystals as it is related to the local field strength. This can
significantly facilitate nonlinear optical mixing in such
structures.

As a logical step forward, we can ask what would happen if a
periodically nanostructured metal film would be embedded in a
nonlinear optical material. In this Letter we report first
experimental studies of nonlinear optical properties of arrays of
artificial nanometer-scale cylindrical pinholes in free standing
thick gold films, which exhibit strong nonlinear optical mixing
due to the presence of nonlinear material inside the pinholes and
on the interfaces of the film. In addition to the SPP effects on
the film interfaces, due to pronounced cylindricality of the
holes, the shape-resonances (cylindrical surface plasmons) can be
excited in the holes \cite{boardman}. At a given transmitted
optical power, for a small cylinder diameters optical field of a
cylindrical surface plasmon grows inversely proportional to the
diameter. Thus, sufficiently thin cylindrical pinholes filled with
nonlinear optical material should start to exhibit enhanced
nonlinear optical mixing.

In our experiments rectangular arrays of nanometric cylindrical
holes have been patterned into a freestanding 400 nm thick gold
membrane using focused gallium ion beam (FIB) milling. To avoid
stress in the gold membrane, the gold film is supported by 50 nm
thick $Si_3N_4$ membrane with 5 nm intermediate Cr layer. Since
there is no glass/metal interface, milling artifacts such as
redeposited material can be avoided leading to a better-defined
structure geometry. We have been successful in patterning
relatively large areas (500 $\mu$m x 500 $\mu$m) of the gold
membranes, creating nanosized holes with dimensions down to 20 nm.
The diameter of the hole can be controlled by using a controlled
endpoint termination technique. In this technique a Faraday cup is
placed underneath the membrane and the transmitted ion current is
measured {\it in-situ} during FIB milling. The milling can then be
terminated, by blanking the ion beam, once a defined ion current
is passing through the emerging hole in the membrane. The images
of the cylindrical pinhole arrays fabricated in a 400 nm thick
free standing gold film are shown in Fig. 1. Similar to the sample
preparation described in \cite{prl}, a drop of
poly-3-butoxy-carbonyl-methyl-urethane (3BCMU) polydiacetylene
solution in chloroform was deposited onto the gold film surface.
After solvent evaporation films of polydiacetylene were left on
both the top and the bottom surfaces of the perforated gold film.
Such 3BCMU polydiacetylene films hold the current record for the
largest fast nonresonant optical $\chi^{(3)}$ nonlinearity
\cite{10}.

\begin{figure}[tbp]
\centerline{\psfig{figure=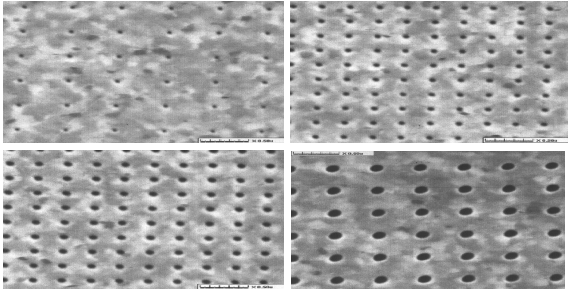,width=9.0cm,height=6.0cm,clip=}}
\caption{Focused ion beam milled arrays of cylindrical pinholes in
a 400 nm thick free standing gold film. The scale bar is 500 nm.}
\label{fig1}
\end{figure}

Our experimental setup is shown in Fig.2. A bent optical  fiber
tip similar to the bent tips often used in near-field optical
microscopes \cite{11} was used to collect the transmitted light.
The fiber tip was positioned above the individual pinhole array in
contact with the thick polymer coating using a far-field optical
microscope and the shear-force distance control system commonly
used in near-field optical setups. \cite{12} Since pinhole arrays
were typically separated by at least 10 $\mu $m distances,
properties of individual arrays were studied in the absence of any
background scattered light. Under far-field microscope control,
$\sim 30 \times 30 \mu m^2$ area of the sample containing the
pinhole array of interest has been illuminated with 488 nm light
of a CW argon ion laser and the 633 nm light of a CW He-Ne laser.
The latter was used as signal light and the former was used as
control light in the optical gating experiments. The samples were
illuminated at the angle of incidence of approximately 45$^o$ so
that the projection of the illuminating light wavevector onto the
surface is in the direction approximately corresponding to the
direction along the hole rows ($\Gamma-X$ direction of the
Brillouin zone).

\begin{figure}[tbp]
\centerline{\psfig{figure=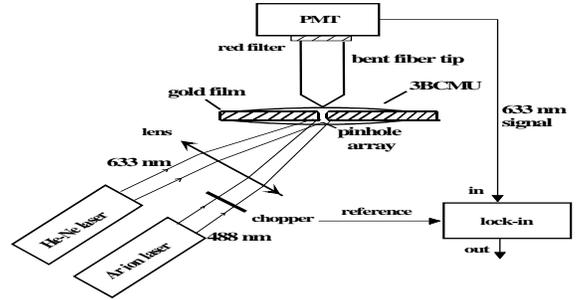,width=9.0cm,height=6.0cm,clip=}}
\caption{ Schematic view of the experimental setup. } \label{fig2}
\end{figure}

First we have studied how transmission of the arrays at the signal
light wavelength ($\lambda$ = 633 nm) is affected by simultaneous
illumination with control light ($\lambda$ = 488 nm). In order to
observe light controlled tunneling directly, we modulated the
incident 488 nm light with a chopper at a frequency of 1.2 kHz.
Light transmitted through the pinhole array was collected with the
fiber and detected with a photomultiplier (PMT) through an optical
filter which completely cuts off the blue light.  Modulation of
the transmission of the array with 20 nm holes at the 633 nm
wavelength induced by modulation of P-polarized control light
illuminating the same area was observed using a lock-in amplifier
,
which was synchronized with the chopper (Fig. 2). Thus, the
intensity variations of the signal light tunneling through the
pinhole array induced by 488 nm light have been measured directly
(Fig. 3). The time behavior of the observed switching is
determined by the time constant of the lock-in-amplifier ($\sim$ 3
s) needed for the signal integration. At the same time, for the
arrays with larger holes ($\sim$ 100 nm diameter) no significant
variations of the transmission was observed for the given
wavelengths and intensities of signal and control light.

\begin{figure}[tbp]
\centerline{\psfig{figure=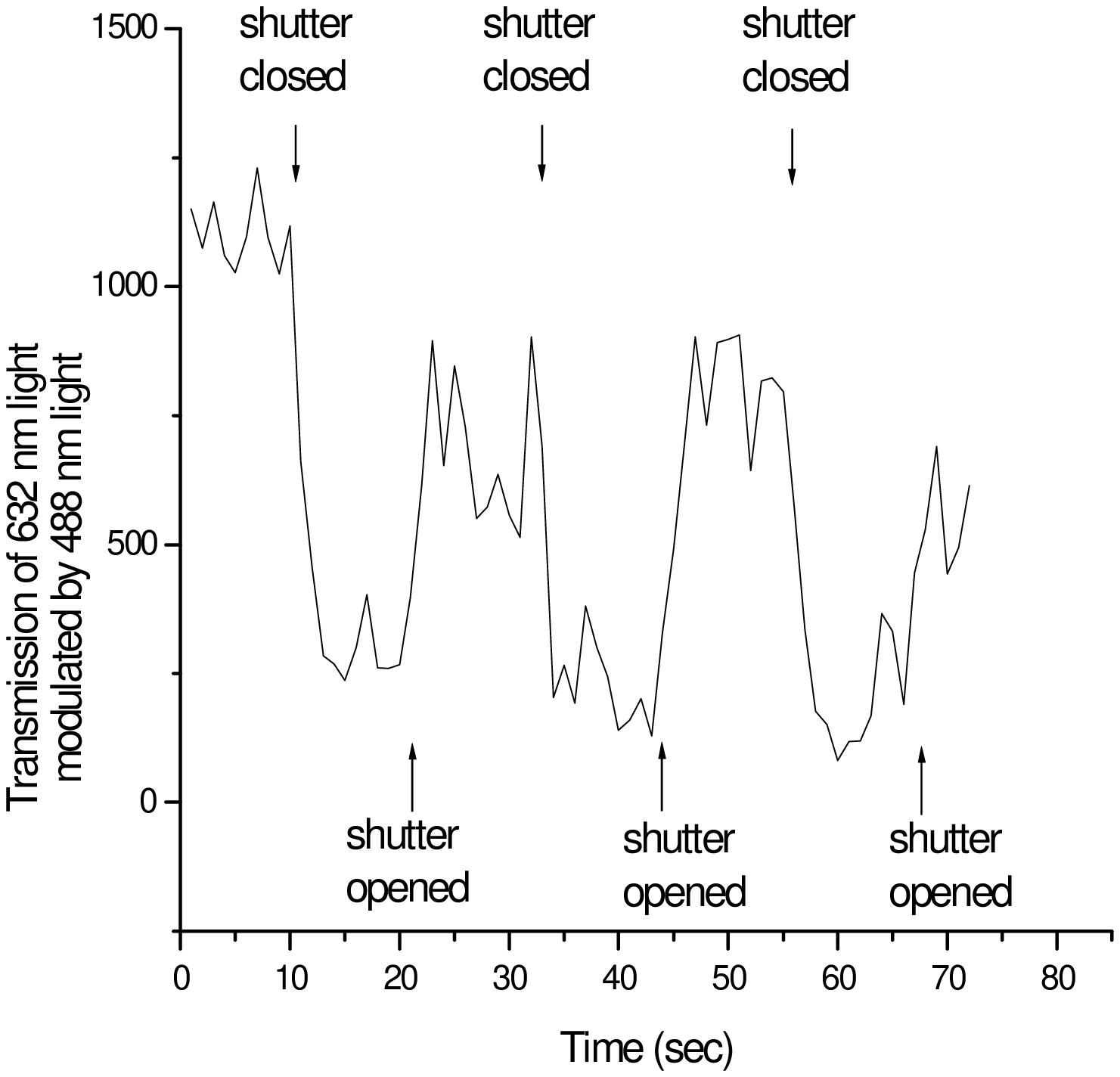,width=9.0cm,height=6.0cm,clip=}}
\caption{Modulation of the 20 nm pinholes array transmission at
633 nm induced by modulation of control light. The shutter of
control light was closed and opened three times during the
experiment.} \label{fig3}
\end{figure}

Polarization properties of the signal light passing through the
array of smallest pinholes are strikingly different from the
polarization properties of the transmitted red light measured in
the absence of blue light illumination. These polarization
properties provide a very strong proof of nonlinear optical mixing
in the polydiacetylene-filled pinholes. In the absence of
modulating blue light, transmission of P-polarized red light is
higher than the transmission of S-polarized red light by
approximately a factor of 3. This observation is consistent with
the theoretical model of the enhanced transmission of
subwavelength pinhole arrays relying on the excitation of surface
polaritons on both film interfaces, since at the oblique incidence
the light of different polarizations interacts with different SPP
resonances. This ratio of P- to S-polarized light transmission has
been observed to be rather insensitive to the pinhole diameter in
a studied range of the pinhole sizes as is expected since the
spectrum of the SPP excitations depends mainly on the periodicity
of the structure.

However, under the modulation with P-polarized blue light, the P
to S ratio in the modulated red light transmission jumps to
approximately a factor of 20. Polarization measurements offer the
best distinction between nonlinear optical mixing effects of
interest and possible thermal effects such as thermal expansion of
the pinholes, heating of the filling material etc. which could
interfere with the nonlinear optical effects. Thermal modulation
of subwavelength aperture transmission has been observed
previously under intense aperture illumination \cite{13}. We were
able to observe similar effects in the pinhole arrays transmission
at much larger than usual illumination intensities achieved by
much tighter focusing (to within a-few-micrometers spots) of the
modulating blue light. However, polarization dependence of the
modulated red light obtained under such conditions is very weak
(Fig.4(c)): the P to S ratio in the thermally modulated red light
transmission falls to approximately a factor of 1.5.

\begin{figure}[tbp]
\centerline{\psfig{figure=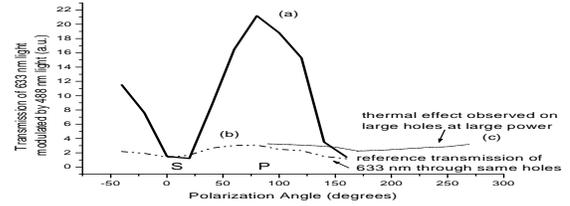,width=9.0cm,height=6.0cm,clip=}}
\caption{ (a) Polarization dependence of the modulation of the 20 nm
pinholes array transmission at 633 nm induced by modulation of
P-polarized 488 nm light. (b) Polarization dependence of the same
array transmission at 633 nm without illumination with blue light.
(c) Polarization dependence of the thermal modulation of the red light
transmission.} \label{fig4}
\end{figure}

Extremely strong polarization dependence of the modulated
transmission of the signal light can be explained taking into
account the properties of surface polariton Bloch waves on the
periodically perforated gold surfaces, the properties of surface
plasmons in cylindrical channels, and the properties of a $\chi
^{(3)}$ nonlinear susceptibility tensor of 3BCMU polydiacetylene.

To understand the observed transmission behavior, let us consider
the spectrum of electromagnetic excitations in our system. At
certain combinations of incidence angles and wavelengths, the
electromagnetic wave incident on the periodic structure excites the
surface polariton Bloch waves. Diffraction of light on a periodic
structure provides the wave vector conservation needed for the
coupling to SPP \cite{raether}:
\begin{equation}
{\mathbf{k}}_{SP}=\frac{\omega}{c}sin\theta {\mathbf{u}}_{xy}
\delta_{s-p} \pm p \frac{2\pi}{D} {\mathbf{u}}_x \pm q
\frac{2\pi}{D} {\mathbf{u}}_y, \label{SPP-res}
\end{equation}
where $\delta_{s-p}$ = 1 for P-polarized incident light (defined
with respect to the film interface) and 0 for S-polarized light,
$\mathbf{k}_{SP}$ is the surface polariton wave vector on a
periodic interface, ${\mathbf{u}}_{xy}$ is the unit vector in the
direction of the in-plane component of the incident light wave
vector, ${\mathbf{u}}_x$ and ${\mathbf{u}}_y$ are the unit
reciprocal lattice vectors of a periodic structure, D is its
periodicity (assumed to be the same in both x- and y-directions),
and p and q are the integer numbers corresponding to the different
propagation directions of the excited SPP modes. Since SPP is in
general a longitudinal excitation, the electric field of the
excitation light should have an electric field component
perpendicular to the surface or parallel to the SPP propagation
direction. Therefore, according to Eq.(\ref{SPP-res}) surface
polaritons excited with S-polarised light correspond to the Bloch
waves at the edge of the even Brillouin zones (standing modes),
while for P-polarized excitation they are propagating Bloch waves.

In our case, the metal film is sufficiently thick to neglect the
interaction between surface polaritons excited on different
interfaces, therefore two independent sets of resonances can be
considered \cite{raether}. Since one side of the metal film is
covered with a polymer ($\epsilon \approx$ 1.7) while another is
in contact with Cr/Si$_3$N$_4$ ($\epsilon \approx$ 2.0), the SPP
resonant conditions will be slightly different on different
interfaces of the film. (The dispersion of surface polaritons on a
smooth surface is given by
$k_{SP}=\omega/c(\epsilon\epsilon_m/(\epsilon+\epsilon_m))^{1/2}$,
where $\epsilon_m$ and $\epsilon$ are the dielectric functions of
metal and adjacent medium, respectively.)  Since the filling
factor of the structures under consideration is small enough
($f=(d/D)^2$) the modifications of the SPP dispersion on a smooth
interface is expected to be also small, and it can be used to
estimate the resonant frequencies of the SPP Bloch waves
excitation. Considering $\Gamma-X$ direction of the Brillouin zone
($p= \pm 1, \pm 2,...;q=0$), for P-polarised incident light and
the angle of incidence of 45$^o$ used in the experiment, the SPP
Bloch waves corresponding to 3rd and 4th Brillouin zone are
excited at the wavelengths close to $\lambda$ = 633 nm on the
polymer-metal interface. The SPP states related to the illuminated
(Si$_3$N$_4$) interface will be red-shifted with respect to this
frequency. At the same time for S-polarised light, the resonant
SPP frequencies on the metal-polymer interface are far away from
the signal light frequency while SPP Bloch waves from the 4th
Brillouin zone of the illuminated interface are relatively close
to it. This SPP is about twice weaker than the SPP on a pure gold
surface due to presence of a Cr layer which has strong losses at
this wavelength ($Im\epsilon\approx$ 30). Analogous picture can be
constructed in the $\Gamma-M$ direction of the SPP Brillouin zone,
but the related resonant frequencies $(p,q)=(\pm 1,\pm 1)$ will
lie between the frequencies determined by $(p,q)=(\pm 1,0)$ and
$(p,q)=(\pm 2,0)$, and therefore, far from the signal wavelength.
However, these resonances can become important for other angles of
incidence.

To complete the picture, the resonances associated with
cylindrical channels in a metal film should be considered
\cite{boardman}. For thick films with well defined cylindricality
of the channels ($d<<h$), the spectrum of surface electromagnetic
excitations in channels has a rather complicated structure with
both radiative and nonradiative modes present. Individual
cylindrical channels will have a discrete spectrum of resonances
asymptotically approaching surface plasmon frequency ($\epsilon _m
= - \epsilon $) from the high frequency side. For polymer-filled
channels these modes overlap the frequency of the control blue
light for large quantum numbers $s>>1$. The interaction between
channels in an array can additionally broaden these resonances
leading to minibands \cite{kuzmiak-94}. Thus, quasi-continuous
spectrum of the states related to the cylindrical surface plasmons
can be expected in the spectral range of the control light. For
infinitely long cylindrical channels the spectrum of wave vectors
along the cylinder axis $h_z$ is continuous. Physically one can
imagine these excitations as surface modes of a spiral trajectory
on a cylinder channel surface. Real (nonradiative) surface modes
can not be excited directly by light, but at the frequency
corresponding to the control light the very long wave vector SPP
can be excited on the periodically perforated polymer-metal
interface, which then can be coupled to cylindrical surface
plasmons.

In the absence of control light, the transmission of red light
takes place via resonant tunneling through the states of SPP Bloch
waves on the polymer-metal interface. The changes of the incident
light polarization result in the shift of the SPP resonances and,
hence, variation of the transmission. Being confined to the
polymer-coated interface, the SPP modes responsible for this
transmission are very sensitive to the dielectric constant of the
polymer, since changes of the dielectric constant modify the SPP
resonant conditions and the transmission coefficient. Control
(blue) light coupled into cylindrical surface plasmons either via
their radiative part or via surface polaritons results in the local
changes of the dielectric constant of the polymer due to
third-order Kerr nonlinearity. Local electromagnetic field is
strongly enhanced in and around the channel due to cylindrical
surface plasmons excitation because of small volume of these
surface modes ($E_L \sim 1/d$).

The polarization properties of the polymer molecules themselves
may play significant role in the observed increase of the P to S
ratio in the modulated transmission of red light. The third-order
nonlinearity of 3BCMU polydiacetylene is contributed mainly by the
$\pi$-electrons in the backbone of the polymer \cite{10}. Each
straight segment of polymer backbone could be treated as identical
one-dimensional rod-like chromophore. At the microscopic level the
$\chi^{(3)}$ tensor of 3BCMU is dominated by only one component:
$\chi^{(3)}_{ssss}$, where $s$ is the direction of the polymer
chain. This fact has been verified in measurements of different
components of the macroscopic $\chi ^{(3)}$ tensor of spin-coated
thin polymer 3BCMU films, where polymer backbones have random
in-plane orientations, determined by a flat substrate.\cite{10} In
our case, the preferential direction for the long  polymer
backbones inside narrow cylindrical channels ought to be the
direction along the channel. The nonlinear optical mixing of
interest is determined by the components $D_l(\omega _2)$ of the
optical field inside the channels at $\lambda_2$ = 633 nm:
\begin{equation}
D_l(\omega _2)=\chi ^{(3)}_{ijkl}E_i(\omega _1)E_j(\omega _1)E_k(\omega _2),
\end{equation}
where $\omega _1$ and $\omega _2$ are the frequencies of blue and
red light, respectively. If the zzzz-components (along the channel
direction) of $\chi ^{(3)}_{ijkl}$ dominate the third-order
susceptibility, the modulated red light must be strongly
P-polarized. At the same time, at the frequency of the blue light
$\omega _1$ the electromagnetic field in the cylindrical channels
is dominated by the plasmon modes with large wave vectors along
the channel, for which optical field oscillations have significant
longitudinal component, which is along the channel direction. Thus,
one should expect the enhanced nonlinear optical mixing to occur
while electromagnetic field is traveling through small
cylindrical holes.

In conclusion, we have reported first observation of photon
tunneling gated by light at a different wavelength in an
artificially created array of nanometer scale cylindrical channels
in a thick gold film. Polarization properties of gated light
provide strong proof of the enhanced nonlinear optical mixing in
nanometric channels involved in the process.

\end{document}